\documentclass[doublecol]{epl2}
\usepackage[T1]{fontenc}
\usepackage[latin1]{inputenc}
\usepackage{amsmath}
\usepackage{graphicx}

\title{Optical properties of graphene: the Fermi liquid approach}

\shorttitle{Optical properties of graphene}

\author{M. I. Katsnelson}
\shortauthor{M. I. Katsnelson}

\institute{Institute for Molecules and Materials, Radboud
University Nijmegen, 6525 ED Nijmegen, The Netherlands
 }




\abstract{Optical properties of two-dimensional massless Dirac
fermions are considered by the formalism of pseudospin precession
equations which provides an easy and natural semiphenomenological
way to include correlation effects. It is shown that the latter
are negligible, with the only assumption that the system under
consideration is normal Fermi liquid. This result probably
explains recent experimental data on the universal optical
conductivity of graphene (Nair R. R. et al, {\it Science}  {\bf
320} (2008) 1308).}

\begin{document}

\maketitle

The recent discovery of the first purely two-dimensional material,
graphene \cite{kostya0,kostya1} and of its peculiar electronic
spectrum with chiral massless charge carriers (``Dirac fermions'')
\cite{kostya2,kim} lead to an explosion of scientific activity
(for review, see Refs. \cite{reviewGK,reviewktsn,reviewRMP}).
Among other unique properties of graphene, its universal optical
conductivity is of special interest. It was demonstrated
experimentally that visual transparency of graphene is determined
only by the fine structure constant \cite{fine} (the same
universal optical conductivity has been observed for graphite
\cite{kuzmenko}). This result is in agreement with the theory for
noninteracting Dirac fermions (see Refs. \cite{gusynin,stauber}
and references therein), within an accuracy of 5\%. However, this
is a problem since, generally speaking, one could expect an
essential many-body renormalization of the optical conductivity.
It has been demonstrated before the discovery of graphene that
Coulomb interactions modify drastically the properties of
two-dimensional Dirac fermions making them a marginal Fermi liquid
with strong logarithmic renormalization of the Fermi velocity and
related properties \cite{nuclphys}. Recent explicit calculations
\cite{herbut,sachdev} result in corrections to the
frequency-dependent conductivity $\sigma(\omega)$ of order of
$1/\ln{\left( W/ \hbar \omega \right)}$ where $W$ is a cutoff
energy of the order of the bandwidth. These corrections are for
sure larger than the experimental errors.

Actually, the relevance of correlation effects in graphene is
still rather controversial. In particular, direct measurements of
electron compressibility in graphene \cite{yacoby} do not find any
essential difference with the predictions of the noninteracting
Fermi-gas model. At the same time, some many-body features were
observed in the infrared conductivity \cite{basov}. Theoretically,
the problem seems to be also rather complicated. For example,
taking into account ripples on graphene and related gauge fields
\cite{solstcomm,phil,guinea} can drastically change the picture of
electron-electron interactions \cite{herbut,aleiner}.

In this situation it seems reasonable to investigate the problem
semi-phenomenologically, in the spirit of the Landau Fermi liquid
theory \cite{landau,pw,vk}. The applicability of this theory to
charge carriers in graphene is unclear now. However, in the
current controversial situation it may be reasonable to start
``from the answer''. It will be shown here that, in the framework
of the Fermi liquid theory, the correlation renormalization of the
optical conductivity are almost cancelled so that the experimental
results \cite{fine} may find a natural explanation. This is not
trivial since, in a standard situation, the Fermi liquid theory
allows to include all correlation effects in the renormalization
of parameters only for {\it static} properties whereas at finite
frequencies essentially many-body effects can be expected
\cite{pw,vk}. Since no alternative explanations are known yet it
may be a serious motivation for a deeper microscopic study which
would allow to justify the Fermi liquid theory for graphene.

We will restrict ourselves to the model of Dirac fermions
neglecting valley and spin degrees of freedom. Then, the effective
Hamiltonian for noninteracting fermions in the presence of a
uniform time-dependent electric field $\mathbf{E}\left( t\right) $
reads
\begin{equation}
H=\sum\limits_{\mathbf{p}}\Psi _{\mathbf{p}}^{\dagger }\left( v\mathbf{%
p\sigma }-ie\mathbf{E}\nabla _{\mathbf{p}}\right) \Psi
_{\mathbf{p}} \label{hamil}
\end{equation}
where $v$ is the electron velocity, $\mathbf{p}$ is the quasimomentum, $%
\mathbf{\sigma }=\left( \sigma _x,\sigma _y\right) $ are the Pauli
matrices,
$e$ is the electron charge, and $\Psi _{\mathbf{p}}^{\dagger }=\left( \psi _{%
\mathbf{p}1}^{\dagger },\psi _{\mathbf{p}2}^{\dagger }\right) $
are electron creation operators depending on the pseudospin
(sublattice) index $i=1,2$; here and further $\hbar =1.$ The
canonical transformation
\begin{eqnarray}
\psi _{\mathbf{p}1} &=&\frac 1{\sqrt{2}}\left( \xi _{\mathbf{p}1}+\xi _{%
\mathbf{p}2}\right) ,  \nonumber \\
\psi _{\mathbf{p}2} &=&\frac{\exp \left( i\phi _{\mathbf{p}}\right) }{\sqrt{2%
}}\left( \xi _{\mathbf{p}1}-\xi _{\mathbf{p}2}\right)
\label{unit}
\end{eqnarray}
introduces the annihilation operators for the hole and electron
states, $\xi
_{\mathbf{p}1},\xi _{\mathbf{p}2}$with the energies $\varepsilon _{\mathbf{p}%
1,2}=\mp vp,$ respectively, $\phi _{\mathbf{p}}$ is the polar
angle of the vector $\mathbf{p}$ (for a detailed discussion of the
Hamiltonian and transformation see, e.g., Ref. \cite{aus}). In the
collisionless limit the
equation of motion for the average density matrix $\rho _{\mathbf{p}}=\Psi _{%
\mathbf{p}}^{\dagger }\Psi _{\mathbf{p}}$ has the form
\begin{equation}
i\frac{\partial \left\langle \rho _{\mathbf{p}}\right\rangle }{\partial t}%
=\left\langle \left[ H,\rho _{\mathbf{p}}\right] \right\rangle =v\mathbf{p}%
\left\langle \left[ \sigma ,\rho _{\mathbf{p}}\right] \right\rangle -ie%
\mathbf{E}\left( t\right) \nabla _{\mathbf{p}}\left\langle \rho _{\mathbf{p}%
}\right\rangle .  \label{liuville}
\end{equation}
Introducing scalar ($n$) and pseudospin ($\mathbf{m}$) densities
by a decomposition
\begin{equation}
\left\langle \rho _{\mathbf{p}}\right\rangle =n_{\mathbf{p}}{\rm I}+\mathbf{m}_{%
\mathbf{p}}\mathbf{\sigma }
\end{equation}
where ${\rm I}$ is the two by two unit matrix, one has a set of
uncoupled equations of motion,
\begin{eqnarray}
i\frac{\partial n_{\mathbf{p}}}{\partial t} &=&-ie\left( \mathbf{E\cdot }\nabla _{\mathbf{p}%
}\right)n_{\mathbf{p}},  \label{n} \\
i\frac{\partial \mathbf{m}_{\mathbf{p}}}{\partial t} &=&2iv\left( \mathbf{%
p\times m}_{\mathbf{p}}\right) -ie\left( \mathbf{E\cdot }\nabla _{\mathbf{p}%
}\right) \mathbf{m}_{\mathbf{p}}.  \label{m}
\end{eqnarray}
Only the second one is relevant for us since the electron current
does not depend on $n_{\mathbf{p}}$:
\begin{equation}
\mathbf{j}=2ev\sum\limits_{\mathbf{p}}\mathbf{m}_{\mathbf{p}}.
\label{j}
\end{equation}
Further we will consider only linear optical effects assuming $\mathbf{E}%
\left( t\right) =\mathbf{E}\exp \left( -i\omega t\right) $ and
using the
linear approximation for $\mathbf{m}_{\mathbf{p}}=\mathbf{m}_{\mathbf{p}%
}^{(0)}+\delta \mathbf{m}_{\mathbf{p}}\exp \left( -i\omega
t\right)$, $\delta \mathbf{m}_{\mathbf{p}}\sim {\rm E}$.

At last, using the unitary transformation (\ref{unit}), we derive
\begin{equation}
\mathbf{m}_{\mathbf{p}}^{(0)}=\frac{\mathbf{p}}{2p}\left( f_{\mathbf{p}1}-f_{%
\mathbf{p}2}\right)   \label{zero}
\end{equation}
where $f_{\mathbf{p}i}=\left\langle \xi _{\mathbf{p}i}^{\dagger }\xi _{%
\mathbf{p}i}\right\rangle $ are the Fermi functions of the
energies $\mp vp$. This vector lies in the $xy$ plane and
therefore $\delta m^z$ is not coupled with the electric field.
Excluding $\delta m^z$ from Eq.(\ref{m}) (further we will omit the
subscript $\mathbf{p}$ for brevity) we obtain
\begin{eqnarray}
\left( \omega ^2-4v^2p_y^2\right) \delta m^x+4v^2p_xp_y\delta m^y
&=&-ie\omega E\frac{\partial m^{x\left( 0\right) }}{\partial p_x},
\nonumber
\\
4v^2p_xp_y\delta m^x+\left( \omega ^2-4v^2p_y^2\right) \delta m^y
&=&-ie\omega E\frac{\partial m^{y\left( 0\right) }}{\partial p_x}
\label{linear} \\
&&  \nonumber
\end{eqnarray}
where we have chosen the direction of $x$ axis along the electric
field. After a straightforward transformations we find $j_x=\sigma
\left( \omega \right) E$ where the proportionality coefficient,
that is, the optical conductivity, equals
\begin{equation}
\sigma \left( \omega \right) =-\frac{8ie^2v^3}\omega \sum\limits_{\mathbf{p}}%
\frac{p_y}{\omega ^2-4v^2p^2}\left( p_y\frac{\partial m^{x\left( 0\right) }}{%
\partial p_x}-p_x\frac{\partial m^{y\left( 0\right) }}{\partial p_x}\right) .
\label{sigma1}
\end{equation}
For the case of zero doping and zero temperature,
$f_{\mathbf{p}1}=1, f_{\mathbf{p}2}=0$ and we have a well-known
result for universal, frequency-independent optical conductivity
$\sigma \left( \omega \right) =e^2/16=\pi e^2/8h$ (per valley per
spin) \cite{fine,kuzmenko,gusynin,stauber}.

This method of derivation can be easily generalized on the case of
\textit{interacting} electrons in the Fermi liquid theory
approach. The only essential difference with the standard case
\cite{pw,vk} is that we have to
work with the matrix distribution function $\left\langle \rho _{\mathbf{p}%
}\right\rangle $.

The interaction effects are taken into account by the replacement
$\left[
H,\delta \rho _{\mathbf{p}}\right] \rightarrow $ $\left[ H,\delta \rho _{%
\mathbf{p}}\right] +$ $\left[ \delta H,\rho _{\mathbf{p}}^{\left(
0\right) }\right] $ in the linearized version of the equation of
motion (\ref{liuville}) where
\begin{equation}
\delta H_{\mathbf{p}}=\sum\limits_{\mathbf{p}^{\prime }}F_{\mathbf{pp}%
^{\prime }}\left \langle \delta \rho _{\mathbf{p}^{\prime }}
\right \rangle \label{landau}
\end{equation}
and $F_{\mathbf{pp}^{\prime }}$ is the (matrix) Landau interaction
function \cite{landau,pw,vk}. For the case under consideration
(zero doping
and zero temperature) $\left \langle \rho _{\mathbf{p}}^{\left( 0\right) }\right \rangle=-\frac{\mathbf{%
p\sigma }}{2p}.$

Now we have to use symmetry considerations to specify the Landau
function. First, $F$ should be rotationally invariant in the
two-dimensional space.
Second, due to inversion and time-reversal symmetry, it cannot contain $%
\sigma ^z$ matrices \cite{sym}. Third, it should vanish at
$\mathbf{p}^{\prime } \rightarrow 0$ or $\mathbf{p} \rightarrow 0$
since electron-electron interactions cannot open the gap without
symmetry breaking \cite{sym}. Thus, we have
\begin{equation}
F_{\mathbf{pp}^{\prime }}=A {\rm I}\otimes {\rm I^{\prime }}+B
\left( \mathbf{p\sigma }\right)\otimes \left( \mathbf{p}^{\prime }\mathbf{\sigma }%
^{\prime }\right)  +C \left(\mathbf{p}\mathbf{p^{\prime }} \right)
\left( \sigma^x \otimes \sigma^{\prime x} + \sigma^y \otimes
\sigma^{\prime y} \right) \label{F}
\end{equation}
where $A$, $B$, and $C$ are some unknown functions of $\left|
\mathbf{p-p}^{\prime }\right|$. In particular, the long-range
Coulomb (Hartree) interaction singular at small momentum transfer
contributes to the $A$ function only. As for the ``exchange''
interactions $B$ and $C$ one can assume that they are smooth
functions which can be expanded in powers of the momentum transfer
square. Substituting Eqs.(\ref{landau}) and (\ref{F}) into the
equation of motion one finds:
\begin{eqnarray}
\omega ^2\delta m^x-4v^2p_y^2 \delta \tilde{m^x} +4v^2p_xp_y
\delta \tilde{m^y}  &=&-ie\omega E\frac{\partial
m^{x\left( 0\right) }}{\partial p_x},  \nonumber \\
4v^2p_xp_y \delta \tilde{m^x} +\omega ^2\delta
m^y-4v^2p_y^2 \delta \tilde{m^y}  &=&-ie\omega E\frac{%
\partial m^{y\left( 0\right) }}{\partial p_x}  \label{linearFL} \\
&&  \nonumber
\end{eqnarray}
where $\delta \tilde{{\bf m}} = \delta {\bf m} + {\bf \Delta}$,
\begin{equation}
\mathbf{\Delta }_{\mathbf{p}}=\frac{1}{vp}\sum\limits_{\mathbf{p}%
^{\prime }} \left[ B_{\mathbf{p}\mathbf{p^{\prime}}} \mathbf{p}
\left( \mathbf{p}^{\prime }\delta \mathbf{m}_{\mathbf{p}^{\prime
}}\right) + C_{{\mathbf{p}\mathbf{p^{\prime}}}} \left( \mathbf{p}
\mathbf{p}^{\prime } \right)\delta \mathbf{m}_{\mathbf{p}^{\prime
}} \right] \label{delta}
\end{equation}
contains all correlation effects. Eq.(\ref{linearFL}) is the
analog of Eq.(\ref{linear}), differing by the terms with $\Delta$.
The latter give an additional contribution to the current density
which can be represented, after simple transformations, in the
form
\begin{equation}
j_x^{corr}=8e^2v^3\sum\limits_{\mathbf{p}}\frac{p_y}{\omega
^2-4v^2p^2}\left( p_y\Delta ^x-p_x\Delta ^y\right)
\label{correction}
\end{equation}
Note that the terms with $B$ function are exactly cancelled in the
expression (\ref{correction}) and only the term proportional to
$C$ can, in general, survive. However, it vanishes obviously by
symmetry if one assume $C = const$.

To find the correlation corrections to the optical conductivity
explicitly in a general case one has to solve the integral
equations (\ref{linearFL}),(\ref{delta}). Fortunately, their
frequency dependence can be found just analyzing perturbation
expansion in the interaction functions $B$ and $C$. One can see
(it is also obvious from physical considerations) that any
adsorption processes require at least one real (and not virtual)
interband transition and, thus, either $p$ or $p^{\prime }$ should
be equal to $\omega /2v$ (imaginary part of the fraction in
Eq.(\ref{correction}) contains the delta-function). The leading
correlation terms just vanish as was discussed above.

The next terms of the expansion of $C$ in $\left(
\mathbf{p-p}^{\prime }\right) ^2$ should be taken into account
which
gives, at least, one more power of $p$ in the integrand. As a result, $%
j_x^{corr}\propto \omega ^3$ which means, in dimensionless units,
$\left( \hbar \omega /W\right) ^3$. Indeed, since the interaction
constant in graphene $e^2/\hbar v$ is of order unity, the energy
cutoff is the only relevant characteristic which enters the
problem. This is smaller than the corrections to the optical
conductivity because of inaccuracy of the Dirac Hamiltonian itself
\cite{stauber}, which are of order of $\left( \hbar \omega
/W\right) ^2$. Thus, the Fermi liquid interaction contributions to
$\sigma \left( \omega \right) $ are really negligible.

To conclude, experimental data \cite{fine}, together with the
present analysis, seem to support the Fermi liquid picture of
charge carriers in graphene, against the marginal Fermi liquid.
The latter, according to the calculations \cite{herbut} predicts
many-body renormalization of the optical conductivity of order of
$2/\ln \left( W/ \hbar \omega \right)$, that is, of order of
unity. At the same time, our consideration is purely
phenomenological and microscopic justification of the Fermi liquid
picture for graphene is required.

\textbf{Acknowledgements.} The work is financially supported by
Stichting voor Fundamenteel Onderzoek der Materie (FOM), the
Netherlands. I am thankful to Andre Geim for inspiring
discussions.

\end{document}